	\documentclass[
		paper=a4,11pt,american,pagesize,%
		abstract=true,DIV=12,headinclude=true,headlines=0,footinclude=true,footlines=-5,twoside=semi,%
	]{scrartcl}
	\def\docclass{koma}
	\def\version{arxiv}

	\def\draftmode{false} 

\usepackage[utf8]{inputenc}
\makeatletter
\usepackage{xifthen}

\newcommand\iflipics[2]{\ifthenelse{\equal{\docclass}{lipics}}{#1}{#2}}
\newcommand\ifkoma[2]{\ifthenelse{\equal{\docclass}{koma}}{#1}{#2}}
\newcommand\ifieee[2]{\ifthenelse{\equal{\docclass}{ieee}}{#1}{#2}}
\newcommand\ifsiam[2]{\ifthenelse{\equal{\docclass}{siam}}{#1}{#2}}
\newcommand\ifmysiam[2]{\ifthenelse{\equal{\docclass}{my-siam}}{#1}{#2}}
\newcommand\ifacm[2]{\ifthenelse{\equal{\docclass}{acm}}{#1}{#2}}
\newcommand\ifdcc[2]{\ifthenelse{\equal{\docclass}{dcc}}{#1}{#2}}
\ifthenelse{ \equal{\docclass}{lipics} \OR \equal{\docclass}{koma} \OR \equal{\docclass}{ieee} \OR \equal{\docclass}{siam} \OR \equal{\docclass}{my-siam} \OR \equal{\docclass}{acm} \OR \equal{\docclass}{dcc} }{
	\PackageInfo{paper}{Building paper with docclass = \docclass} 
}{
	\PackageWarning{paper}{docclass = "\docclass", but must be one of "lipics", "koma", "ieee", "siam", "my-siam", "acm", "dcc"}
}

\newcommand\ifmanuscript[2]{\ifthenelse{\equal{\version}{manuscript}}{#1}{#2}}
\newcommand\ifarxiv[2]{\ifthenelse{\equal{\version}{arxiv}}{#1}{#2}}
\newcommand\ifsubmission[2]{\ifthenelse{\equal{\version}{submission}}{#1}{#2}}
\newcommand\ifproceedings[2]{\ifthenelse{\equal{\version}{proceedings}}{#1}{#2}}
\ifthenelse{ 
	\equal{\version}{manuscript} 
	\OR \equal{\version}{arxiv} 
	\OR \equal{\version}{submission} 
	\OR \equal{\version}{proceedings} 
}{
	\PackageInfo{paper}{Building paper version = \version} 
}{
	\PackageWarning{paper}{version = "\version", but must be one of "manuscript", "arxiv", "submission", "proceedings"}
}

\newcommand\ifdraft[2]{\ifthenelse{\equal{\draftmode}{true}}{#1}{#2}}
\ifthenelse{ \equal{\draftmode}{true} \OR \equal{\draftmode}{false} }{
	\PackageInfo{paper}{Building paper with draftmode = \draftmode} 
}{
	\PackageWarning{paper}{draftmode = "\draftmode", but must be "true" or "false"}
}

\usepackage[T1]{fontenc}
\ifsiam{
	\usepackage{lmodern}
	\usepackage{slantsc}
}{}
\ifmysiam{
	\usepackage{lmodern}
	\usepackage{slantsc}
}{}
\ifkoma{
	\usepackage{lmodern}
	\usepackage{slantsc}
}{}
\iflipics{
	\usepackage{lmodern}
	\usepackage{slantsc}
}{}
\ifdcc{
	\usepackage{lmodern}
	\usepackage{slantsc}
}{}

\usepackage{babel}
\input{ushyphex.tex} 

\usepackage{array,multicol}
\ifieee{
	\usepackage[cmex10]{amsmath,mathtools}
	\usepackage{amsfonts,amssymb}
}{}
\ifkoma{
	\usepackage{amsmath,amsfonts,amssymb,mathtools}
}{}
\iflipics{
	\usepackage{amsmath,amsfonts,amssymb,mathtools}
}{}
\ifsiam{
	\usepackage{amsmath,amsfonts,amssymb,mathtools}
}{}
\ifmysiam{
	\usepackage{amsmath,amsfonts,amssymb,mathtools}
}{}
\ifacm{
	\usepackage{mathtools}
}{}
\ifdcc{
	\usepackage{amsmath,amsfonts,amssymb,mathtools}
}{}

\usepackage{mleftright}\mleftright 
\usepackage{relsize,xspace,booktabs,adjustbox,needspace,pbox,relsize}
\ifieee{
	
	\usepackage{enumitem}
}{
	\usepackage{enumitem}
}
\usepackage{graphicx}

\ifacm{}{
	\usepackage{colonequals}
}

\usepackage{wref}
\usepackage[bibtex]{url-doi-arxiv}

\ifsiam{
	\usepackage{ltexpprt}
}{}

\newdimen\makeboxdimen
\newcommand\makeboxlike[3][l]{%
\setbox0=\hbox{#2}%
\global\makeboxdimen=\wd0%
\setbox1=\hbox{\makebox[\makeboxdimen][#1]{%
\makebox[0pt][#1]{#3}%
}}%
\ht1=\ht0%
\dp1=\dp0%
\box1%
}

\newcommand\plaincenter[1]{%
	\mbox{}\hfill#1\hfill\mbox{}%
}

\ifthenelse{\equal{\docclass}{koma} \OR \equal{\docclass}{my-siam}}{
	\setlength\parindent{1.5em}
	\usepackage[headsepline]{scrlayer-scrpage}
	\pagestyle{scrheadings}
	\clearscrheadfoot
	\AtBeginDocument{%
		\automark[section]{}%
	}
	\ohead{\pagemark}
	\rehead{\mytitle}
	\lohead{\headmark}
	\addtokomafont{caption}{\sffamily\small}
	\addtokomafont{captionlabel}{\sffamily\textbf}
	\setcapmargin{2em}
}{}
\ifthenelse{\equal{\docclass}{my-siam}}{
	\setcapmargin{1em}
	\setcapindent{0em}
}{}
\ifmysiam{
	\setlength\parskip{0pt}
	\RedeclareSectionCommand[
		beforeskip=-1.25\baselineskip,
		afterskip=0.75\baselineskip,
	]{section}
	\RedeclareSectionCommand[
		beforeskip=-1\baselineskip,
		afterskip=-1.5em,
	]{subsection}
	\RedeclareSectionCommand[
		beforeskip=-1\baselineskip,
		afterskip=-1.5em,
	]{subsubsection}
	\RedeclareSectionCommand[
		beforeskip=-.25\baselineskip,
		indent=1.5em,
		afterskip=-1em,
	]{paragraph}
}{}
\ifdcc{
	\ifproceedings{}{
		\pagestyle{plain}
		\setlength{\footskip}{5ex}
	}
}{}

\AtBeginDocument{%
	\let\mytitle\@title%
}

\newcommand\shorttitle[1]{%
	\ifacm{%
		\FAIL
	}
	\iflipics{%
		\titlerunning{#1}%
	}{}%
	\AtBeginDocument{%
		\def\mytitle{#1}%
	}%
}

\let\oldthebibliography\thebibliography
\renewcommand\thebibliography[1]{%
	\oldthebibliography{#1}%
	\pdfbookmark[1]{References}{}%
}

\usepackage{lscape} 

\ifkoma{
	\usepackage{float}
	\floatstyle{plain}
	\usepackage{newfloat}
	\DeclareFloatingEnvironment[%
			name=Algorithm,%
			placement=thb,%
		]{algorithm}
}{}
\iflipics{
	\usepackage{newfloat}
	\DeclareFloatingEnvironment[%
			name=Algorithm,%
			placement=thb,%
		]{algorithm}
}{}
\ifmysiam{
	\usepackage{newfloat}
	\DeclareFloatingEnvironment[%
			name=Algorithm,%
			placement=thb,%

		]{algorithm}
}

\ifdcc{
	\usepackage{float}
	\usepackage[font={small},labelfont=bf]{caption}
}{}

\ifmysiam{
	\setcounter{topnumber}{3}
	\setcounter{bottomnumber}{3}
	\setcounter{totalnumber}{3}     
	\setcounter{dbltopnumber}{3}    

}{}
\ifsiam{

	\setcounter{topnumber}{3}
	\setcounter{bottomnumber}{3}
	\setcounter{totalnumber}{3}     
	\setcounter{dbltopnumber}{3}    

}{}

\usepackage{dcolumn}

\usepackage{textcomp} 
\usepackage{listings}

\usepackage{tikz}

\usetikzlibrary{positioning,arrows.meta,fit}
\usetikzlibrary{backgrounds,calc,trees,graphs}
\usetikzlibrary{shapes.geometric,shapes.misc}

\pgfdeclarelayer{background}
\pgfsetlayers{background,main}

\iflipics{
	\newtheorem{fact}[theorem]{Fact}

	\newenvironment{proofof}[1]{%
		\begin{proof}[{{Proof of #1{}}}]%
	}{%
		\end{proof}%
	}
}{}
\ifacm{
	\AtEndPreamble{
		\theoremstyle{acmdefinition}
		\newtheorem{remark}[theorem]{Remark}
		\newtheorem{fact}[theorem]{Fact}
	}
	
	\newenvironment{proofof}[1]{%
		\begin{proof}[{{Proof of #1{}}}]%
	}{%
		\end{proof}%
	}
}{}
\ifsiam{
	\newtheorem{remark}{Remark}
	\newenvironment{proofof}[1]{%
			\begin{proof}[{{#1{}}}]%
		}{%
			\end{proof}%
		}
}{}
\ifthenelse{\equal{\docclass}{lipics} \OR \equal{\docclass}{siam} \OR \equal{\docclass}{acm}}{}{
	\usepackage[amsmath,hyperref,thmmarks]{ntheorem}
	
	\theorembodyfont{\slshape}
	\theoremseparator{:}
	\newtheoremstyle{proofstyle}%
	  {\item[\theorem@headerfont\hskip\labelsep ##1\theorem@separator]}%
	  {\item[\theorem@headerfont\hskip\labelsep ##3\theorem@separator]}
	
	\theorempreskip{\topsep} 

	\theoremsymbol{\adjustbox{scale=.8}{$\triangleleft\mkern-1mu$}}
	
	\theoremstyle{plain}
	\theorempreskip{\topsep}

	\theoremstyle{plain}
	\theorembodyfont{\upshape}

	\theoremsymbol{\raisebox{-.25ex}{$\Box$}}
	\qedsymbol{\raisebox{-.25ex}{$\Box$}}
	
	\theoremstyle{proofstyle}
	\newtheorem{proof}{Proof}

}

\iflipics{
	\newenvironment{thmenumerate}[2][]{%
		\begin{enumerate}[
			label={\textsf{\textbf{\color{darkgray}{\makebox[\widthof{(a)}][c]{\textup{(\alph*)}}}}}},
			ref={\ref{#2}\kern.1em--\kern.1em(\alph*)},
			itemsep=0pt,
			topsep=.5ex,
			leftmargin=1.75em,
			#1
		]%
	}{%
		\end{enumerate}%
	}
}{
	
}

\newcommand*\ie{\mbox{i.\hspace{.2ex}e.}}
\newcommand*\eg{\mbox{e.\hspace{.2ex}g.}}

\newcommand*\wrt{\mbox{w.\hspace{.2ex}r.\hspace{.2ex}t.}\xspace}

\usepackage{fixmath}

\newcommand{\ESymbol}{\mathbb{E}}

\newcommand{\ProbSymbol}{\ensuremath{\mathbb{P}}}

\DeclarePairedDelimiterXPP\Prob[1]{\ProbSymbol}[]{}{%
	#1%
}
\DeclarePairedDelimiterXPP\E[1]{\ESymbol}[]{}{%
	#1%
}
\DeclarePairedDelimiterXPP\Eover[2]{\ESymbol_{#1}}[]{}{%
	#2%
}
\DeclarePairedDelimiterXPP\ProbIn[2]{\ProbSymbol_{#1}}[]{}{%
	#2%
}
\providecommand{\Prob}{} 
\providecommand{\ProbIn}{} 
\providecommand{\E}{} 
\providecommand{\Eover}{} 

\iflipics{}{
	\makeatletter
	\let\oldalign\align
	\let\endoldalign\endalign
	\renewenvironment{align}{%
		\begingroup%
		\let\oldhalign\halign
		\def\halign{%
			\let\oldbreak\\%
			\def\nonnumberbreak{\nonumber\oldbreak*}%
			\def\\{%
				\@ifstar{\nonnumberbreak}{\oldbreak}%
			}%
			\oldhalign%
		}
		\oldalign%
	}{%
		\endoldalign%
		\endgroup%
	}
}
\newcommand*\numberthis[1][]{\stepcounter{equation}\tag{\theequation}}

\allowdisplaybreaks[3]

\newcommand\splitaftercomma[1]{%
  \begingroup
  \begingroup\lccode`~=`, \lowercase{\endgroup
    \edef~{\mathchar\the\mathcode`, \penalty0 \noexpand\hspace{0pt plus .25em}}%
  }\mathcode`,="8000 #1%
  \endgroup
}

\def\mydots{\xleaders\hbox to.5em{\hfill.\hfill}\hfill}
\newlength\tmpLenNotations

\ifdraft{%
    \usepackage[switch]{lineno} 
    \linenumbers
	\overfullrule=6mm
	
	\usepackage[color,notref,notcite]{showkeys}
	\definecolor{refkey}{gray}{.99}
	\colorlet{labelkey}{green!60!black!60}
	
	\usepackage[inline,nolabel]{showlabels}

	\showlabels{cite}
	\showlabels{citealt}
	\showlabels{citealp}
	\showlabels{citet}
	\showlabels{Citet}
	\showlabels{citep}
	\showlabels{citeauthor}
	\showlabels{Citeauthor}
	\showlabels{citefullauthor}
	\showlabels{citeyear}
	\showlabels{citeyearpar}
	\showlabels{wref}
	\showlabels{wpref}
	\showlabels{wtpref}
	\showlabels{wildref}
	\showlabels{wildpageref}
	\showlabels{wildtpageref}
}{}

\iflipics{
	\ifmanuscript{\hideLIPIcs}{}
	\ifarxiv{\hideLIPIcs}{}
	\ifsubmission{}{\nolinenumbers}
}{}

\ifdraft{}{%
	\usepackage{microtype}
}

\hypersetup{
	final,
	unicode=true, 
	bookmarks=true,
	bookmarksnumbered=true,
	bookmarksdepth=2,
	bookmarksopen=true,
	breaklinks=true,
	hidelinks,
}

\newsavebox\tmpbox

\ifdcc{
	\renewcommand\paragraph{\@startsection{paragraph}{4}{\parindent}
	                                      {\smallskipamount}
	                                      {-1em}%
	                                      {\normalfont\normalsize\bfseries}}
}{}

\iflipics{
	\let\oldparagraph\paragraph
	\renewcommand\paragraph[1]{%
		\oldparagraph*{#1}
	}
}{
	\let\oldparagraph\paragraph
	\renewcommand\paragraph[1]{%
		\oldparagraph{#1.}
	}
}

\ifmysiam{
	\let\oldsubsection\subsection
	\renewcommand\subsection[1]{%
		\oldsubsection{#1.}%
	}
	\let\oldsubsubsection\subsubsection
	\renewcommand\subsubsection[1]{%
		\oldsubsubsection{#1.}%
	}
}{}
\ifsiam{
	\let\oldsubsection\subsection
	\renewcommand\subsection[1]{%
		\oldsubsection{#1.}%
	}
	\let\oldsubsubsection\subsubsection
	\renewcommand\subsubsection[1]{%
		\oldsubsubsection{#1.}%
	}
}{}

\let\epsilon\varepsilon

\def\myacknowledgements{}
\ifkoma{
	
}{}
\ifieee{
	
}{}
\ifsiam{
	
}{}
\ifmysiam{
	
}{}
\ifdcc{
	
}{}
\ifacm{
	
}{}

\newcommand\pair[2]{\genfrac{[}{]}{0pt}{1}{#1}{#2}}
\newcommand\terminalPair[2]{\pair{\texttt{#1}}{\texttt{#2}}}
\newcommand\ao{\terminalPair{A}{(}}
\newcommand\co{\terminalPair{C}{(}}
\newcommand\go{\terminalPair{G}{(}}
\newcommand\uo{\terminalPair{U}{(}}
\newcommand\ac{\terminalPair{A}{)}}
\newcommand\cc{\terminalPair{C}{)}}
\newcommand\gc{\terminalPair{G}{)}}
\newcommand\uc{\terminalPair{U}{)}}
\newcommand\au{\terminalPair{A}{\textbullet}}
\newcommand\cu{\terminalPair{C}{\textbullet}}
\newcommand\gu{\terminalPair{G}{\textbullet}}
\newcommand\uu{\terminalPair{U}{\textbullet}}

\newcommand\secondaryStructureSymbol[1]{\makeboxlike{\texttt{M}}{\texttt{\bfseries #1}}\xspace}
\newcommand\ssu{\secondaryStructureSymbol{\textbullet}} 
\newcommand\sso{\secondaryStructureSymbol{(}} 
\newcommand\ssc{\secondaryStructureSymbol{)}} 

\usepackage{placeins} 

\makeatother

\usepackage{hyperref}

\ifacm{
	\title[Short Title]{RNA secondary structures: from ab initio prediction to better compression, and back}
}{
	\title{RNA secondary structures: from ab initio prediction to better compression, and back}
	\shorttitle{RNA secondary structures: from prediction to better compression} 
}
\ifdcc{
	\title{\large\bfseries RNA secondary structures: from ab initio prediction to better compression, and back%
	\thanks{%
		Evarista Onokpasa has been supported by the Petroleum Technology Development Fund of Nigeria.
		This work has further been supported in part by the NeST
		(Network Sciences and Technologies) EEECS School initiative of University of Liverpool.
	}}
}{
    \title{RNA secondary structures: from ab initio prediction to better compression, and back}
}

\ifdcc{
	\author{Evarista Onokpasa$^\ast$, Sebastian Wild$^\ast$, and Prudence W.H.\ Wong$^\ast$\\[.5em]
	\begin{minipage}{\linewidth}
	\plaincenter{\begin{tabular}{c}
	$^\ast$University of Liverpool, UK \\ 
	\texttt{\{evarista.onokpasa,\,sebastian.wild,\,pwong\}\,@\,liverpool.ac.uk}\\
	\end{tabular}}
	\end{minipage}%
	}
}{}
\ifkoma{
    \author{
        Evarista Onokpasa%
    \thanks{%
        University of Liverpool, UK,
        \texttt{\{evarista.onokpasa,\,sebastian.wild,\,pwong\}\,@\,liverpool.ac.uk}
    }
    \and
        Sebastian Wild$^\ast$
    \and
        Prudence W.H.\ Wong$^\ast$
    }
}{}

\ifacm{
}{}
\ifthenelse{\NOT \equal{\docclass}{lipics} \AND \NOT \equal{\docclass}{acm} \AND \NOT \equal{\docclass}{dcc}}{
	
	\date{\small\today}
}{}
\ifsiam{
}{}

\begin{document}

\ifacm{}{\maketitle} %

\begin{abstract}
In this paper, we use the biological domain knowledge incorporated into stochastic models for 
\emph{ab initio} RNA secondary-structure prediction to improve the state of the art in 
joint compression of RNA sequence and structure data (Liu et al., BMC Bioinformatics, 2008).
Moreover, we show that, conversely, compression ratio can serve as a cheap and 
robust proxy for comparing the prediction quality of different stochastic models,
which may help guide the search for better RNA structure prediction models.

Our results build on expert stochastic context-free grammar models of RNA secondary structures 
(Dowell \& Eddy, BMC Bioinformatics, 2004; Nebel \& Scheid, Theory in Biosciences, 2011) 
combined with different (static and adaptive) models for rule probabilities
and arithmetic coding.
We provide a prototype implementation and an extensive empirical evaluation,
where we illustrate how grammar features and probability models affect compression ratios.
\end{abstract}

\ifacm{%
	\maketitle%
}{}

\section{Introduction}

In this article, we explore the interplay and potential symbiosis between data compression and 
probabilistic methods for predicting the folding structure of (non-coding) RNA molecules.
Ribonucleic acid (RNA) is a bio-polymer that serves various roles in the coding, decoding, expression and regulation of genes in cells. 
An RNA molecule consists of a chain of \emph{nucleotides} each having a \emph{base} attached to it (either adenine (\texttt A), cytosine (\texttt C), guanine (\texttt G), or uracil (\texttt U)); this string of bases forms the \emph{sequence} of the molecule.
Unlike the related DNA, RNA is usually single-stranded and forms spatial structures by folding onto itself (similar to proteins), with complementary bases forming a stabilizing hydrogen bond.
The set of (indices of the) bases that form such pairs is the \emph{secondary structure} of the molecule;
it can be encoded by the dot-bracket notation,
(see \wref{fig:rna-example}; a formal definition is given in \wref{sec:preliminaries}).

\begin{figure}[tbp]
	\bigskip
	\begin{minipage}[t]{.35\textwidth}
		~\\[-\baselineskip]
		\resizebox{.95\linewidth}!{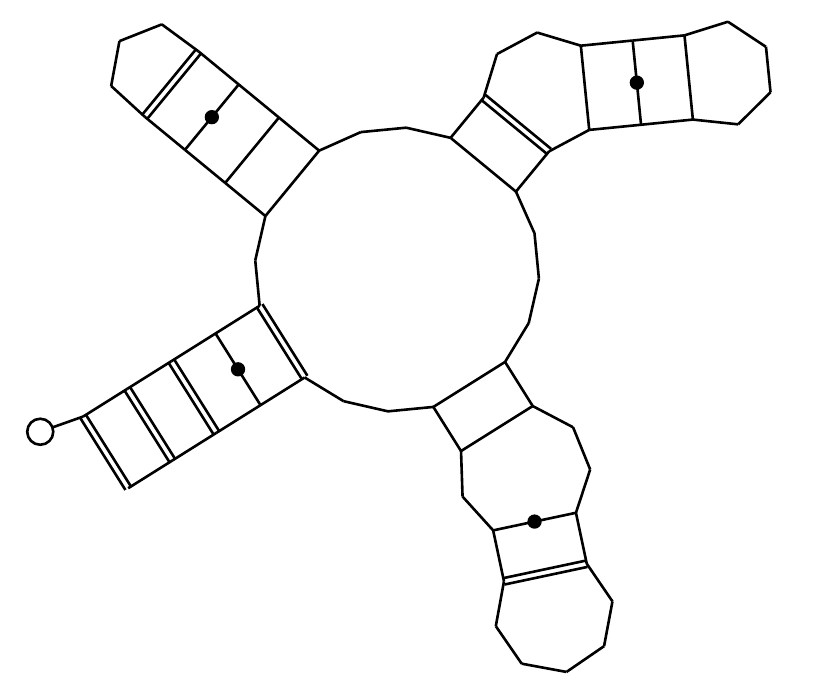}
	\end{minipage}%
	\begin{minipage}[t]{.65\textwidth}
		~\\[-3ex]
		\hspace*{-2em}\resizebox{1.1\linewidth}!{\begin{tikzpicture}[yscale=.2,xscale=.18]
			\foreach \x in {1,...,62} { \node[circle,draw=black,fill=black,inner sep=.8pt] (b\x) at (\x,0) {} ; }
			\foreach \f/\t/\h in {%
				2/62/9,3/61/8.5,4/60/8,5/59/7.5,6/58/7,%
				8/18/3.5,9/17/3,10/16/2.5,11/15/2,%
				21/36/4.5,22/35/4,25/34/3,26/33/2.5,27/32/2,%
				40/55/4,41/54/3.5,44/52/2.5,45/51/2
			} 
			{
				\draw[semithick] (b\f) .. controls ($(b\f)+(0,\h)$) and ($(b\t)+(0,\h)$) .. (b\t) ; 
				\path (b\f) ++(0,-4.2) node[anchor=base] {\scriptsize\sso} ;
				\path (b\t) ++(0,-4.2) node[anchor=base] {\scriptsize\ssc} ;
			}
			\foreach \u in {1,7,12,13,14,19,20,29,23,24,28,29,30,31,37,38,39,42,43,46,47,48,49,50,53,56,57} {
				\path (b\u) ++(0,-4.2) node[anchor=base] {\scriptsize\ssu} ;
			}
			\draw (b1) -- (b62);
			\foreach \i in {1,5,10,...,62,62} {
				\node[scale=.45] at ($(b\i)+(0,-1)$) {\i} ;
			}
			\foreach [count=\i] \b in {G,C,C,C,U,G,A,U,A,G,C,G,U,A,G,U,U,A,C,U,A,G,C,G,A,G,U,C,U,G,U,A,U,U,C,U,A,A,G,A,A,G,A,U,C,A,C,U,G,A,G,G,G,U,U,C,G,C,G,G,G,G} {
				\node[scale=.5] at ($(b\i)+(0,-2.6)$) {\texttt{\b}} ;
			}
			
		\end{tikzpicture}}
		\vspace*{-3ex}
		\caption{%
			An example RNA sequence and structure.
			\textbf{Left:} schematic drawing of structure. 
			\textbf{Above:} Representation as dot-bracket sequence when the backbone is ``pulled straight''.
		}%
		\label{fig:rna-example}
	\end{minipage}
	\vspace*{-2ex}
\end{figure}

The secondary structure is instrumental for the biological function of non-coding RNA molecules and of great interest to biologists.
Much research has hence been devoted to computationally \emph{predict} the secondary structure 
from a known RNA sequence (\emph{ab initio} RNA secondary-structure prediction)%
~\cite{DurbinEddyKroghMitchison1998,GorodkinRuzzo2014,TurnerMathews2016},
including human swarm intelligence~\cite{LeeEtAl2014}, and it
remains an active research area~\cite{SatoAkiyamaSakakibara2021,FuCaoWuPengNieXie2021,SatoKato2021}.
We explore areas around RNA secondary structures where innovations in compression methods
are central for further progress.

\paragraph{Better RNA Compression}
Our first goal is to use the domain knowledge on RNA foldings incorporated into 
secondary-structure prediction models for improved methods for the 
joint compression of the sequence and secondary structure of RNA sequences.
With biological databases ever increasing, compressed representations become desirable.
In the case of databases for non-coding RNA sequences with known secondary structures, 
the data volume has long remained manageable, but growth is now accelerating:
For example, \emph{RNA Central}~\cite{RNACentral2020} now aggregates over 25 million trusted secondary structures 8 years after its first release;
1.8 million of these come from the \emph{rfam} database~\cite{KalvariEtAl2020},
collected over its 20 years of existence.

The need for space-efficient representations of joint RNA sequence and secondary structure databases
has been identified by Liu et al.\ in 2008~\cite{LiuYangChenBuZhangYe2008}. 
Their algorithm \emph{RNACompress}, based on a stochastic context-free grammar (SCFG, defined below), 
has been recognized as an early application 
of ideas from grammar-based compression in the data-compression community~\cite{Maneth2018,KiefferYang2022}.
As we demonstrate in this article, substantially better compression ratios can be achieved than 
Liu et al.\ report; interestingly, by carefully extending their very method to a general 
framework of SCFG-based compression. 
Improvements are then realized by applying this framework on tried and tested grammars
from the RNA secondary structure \emph{prediction} literature~\cite{DowellEddy2004,NebelScheid2011} 
(as well as further, orthogonal refinements).

Apart from the practical utility of less space, compression methods are
of direct interest in bioinformatics as a way to upper bound the Kolmogorov complexity~\cite{Kolmogorov1998} 
of a dataset, and hence its inherent information content~\cite{GiancarloScaturroUtro2009}.
For example in the context of RNA sequences, one can ask how much additional information 
is contained in the secondary structure of the RNA when the sequence is known.

\paragraph{Compression as a proxy for predictive power}

Our second and main goal is to test our hypothesis that for comparing probabilistic models for RNA secondary structures,
\textsl{compression ratio can serve as proxy for prediction quality in RNA secondary-structure prediction}.
Advances in next-generation sequencing allows determining the sequence of many molecules at scale,
whereas secondary structures need to be determined by much more expensive techniques like X-ray crystallography~\cite{TurnerMathews2016}.
A much cheaper and faster alternative is to computationally \emph{predict} the structure from a known sequence.
The state-of-the-art approaches either build on a chemical model of the molecules and try to identify a structure with minimal free energy or use a machine-learning approach.
Both can formally be described by stochastic context-free grammars
(see \wref{sec:preliminaries}).

RNA secondary-structure prediction plays a vital role in studying the biological function of RNA molecules
and for designing artificial RNA sequences, and so numerous software packages implement different algorithms for this task.
Comparing their prediction quality is a delicate undertaking, because no definitive similarity metric is known to judge how close the predicted secondary structure is from an experimentally determined one~\cite{Mathews2019}.
Indeed, the method of choice in the literature to compare structure prediction is solely based on individual base pairs~\cite{Mathews2019,DowellEddy2004,RivasLangEddy2011,NebelScheid2011}:
One compares the \emph{sensitivity} and \emph{positive predictive value (PPV)} of different approaches
(defined in \wref{sec:preliminaries}).

We will use the \emph{compressed size} (in bits per base) of the reference structure under the trained stochastic model
as a more direct means to compare how well different models capture RNA folding behavior.
This compressed size effectively reflects the \emph{log-likelihood} of the reference structure
and hence has a natural interpretation as the information content that model assigns to the RNA structure.

This has several advantages over sensitivity/PPV: 
(a) It directly evaluates the quality of the \emph{model}, separating it from the method to produce a (single) predicted secondary structure.
There are different options to predict a structure; one can use the most likely structure, or a consensus structure containing the most likely individual pairs, or return a sample of several nearly optimal structures. 
No choice clearly dominates the others, but they affect the sensitivity and PPV scores.
(b) Log-likelihood is a single natural metric derived from first principles of information theory; 
it does not need trade-offs or further parameters.

\paragraph{Contributions}

Our contributions are as follows.
First, we improve the compression ratio achieved for joint RNA sequence and structure
data by 45\% over the state of the art, Liu et al.'s RNACompress~\cite{LiuYangChenBuZhangYe2008};
compared to the general-purpose compressor paq8l (\url{http://mattmahoney.net/dc/\#paq}),
we see a 70\% improvement.
The improvement over RNACompress is the combined result of several refinements, but 
a 30\% reduction in compressed size is observed when keeping everything but the used SCFG constant.
This clearly shows the relevance of the grammar and the validity of our approach
to employ structure-prediction grammars.
The proposal and implementation of the more sophisticated grammars 
(such as the one based on~\cite{NebelScheid2011})
is hence a useful contribution. 
Second, we demonstrate that compression ratio can be used as a robust predictor
of how well a grammar will perform for \textit{ab initio} secondary-structure prediction.
To our knowledge, this is the first such attempt to identify suitable probabilistic models 
for RNA structure prediction that is not based on comparing predicted structures 
to a benchmark dataset.
Finally, we reproduce and confirm the computational study of~\cite{DowellEddy2004} with an independent
implementation and additional modifications to their grammars.

\paragraph{Related Work}
\label{sec:related-work}

Liu et al.~\cite{LiuYangChenBuZhangYe2008} proposed RNACompress in 2008;
we discuss their methodology in detail in \wref{sec:rna-compression-scfg}.
Naganuma et al.~\cite{NaganumaHendrianYoshinakaShinoharaKobayashi2020}
explore a related method of SCFG compression 
closer to grammar-based compression using straight-line programs. 
They create a stochastic grammar from the text to compress with a variation of the RePair heuristic~\cite{LarssonMoffat2000}.
For a broader context of grammar-based compression, see the recent survey of Kieffer and Yang~\cite{KiefferYang2022}.
Friemel~\cite{Friemel2020} also targets the joint RNA compression problem,
but using a different approach.
He encodes RNA structures as labeled trees with each
node representing a nucleotide and the branches representing the bonds;
unpaired bases yield unary nodes.
Friemel's algorithm RNAContract contracts sequences of unary nodes 
(similar to compact tries) or a sequence of multiple nested brackets
in the dot-bracket notation. After the node contraction the algorithm
encodes the contracted node tree using Huffman coding.

\paragraph{Outline}
The rest of this paper is structured as follows.
\wref{sec:preliminaries} collects required concepts.
\wref{sec:rna-compression-scfg} explains the grammar-based compression
of RNA.
Then we report on our two studies:
\wref{sec:compression} discusses the compression achieved with various grammars and
\wref{sec:compression-vs-preduction} explores the connection between compressed size and prediction quality.
We conclude in \wref{sec:conclusion} with future work.
In the appendix, we give details on the comparison with a general-purpose compressor (\wref{app:paq8l}),
list the precise grammars we used (\wref{app:grammars}), and investigate further differences between our approach and~\cite{LiuYangChenBuZhangYe2008} (\wref{app:further-results}).
Further details, all datasets and code to produce the figures in this article are available
online as supplementary material: \url{https://www.wild-inter.net/publications/onokpasa-wild-wong-2023};
the code is available on GitHub: \url{https://github.com/evita35/joint-rna-compression}.

\section{Preliminaries}
\label{sec:preliminaries}

\paragraph{Dot-bracket notation}
An RNA sequence is a string of bases \texttt{A}, \texttt{C}, \texttt{G}, \texttt{U}.
Stable hydrogen bonds are possible between \texttt{A} and \texttt{U} resp.~\texttt{C} and \texttt{G}
(the Watson-Crick pairs) and to a lesser extent also between \texttt{G} and \texttt{U}.
RNA secondary structures%
\footnote{%
	As is often done in the area, we do not consider structures with pseudoknots in this paper.
}
can be represented by the dot-bracket notation~\cite{Hofacker1994}:
a well-nested string over $\{\ssu,\sso,\ssc\}$ where a base pair is denoted by matching parentheses
$\sso\ssc$ and an unpaired base by $\ssu$;
see \wref{fig:rna-example} for an example.
We use ``RNA'' as an abbreviation for ``a pair of an RNA sequence and its secondary structure''.

\paragraph{SCFG}
Dot-bracket strings can be generated by a context-free grammar (CFG).
A CFG is a tuple $(N,T,R,S)$ where
$N$ and $T$ are finite sets of \emph{nonterminals} and \emph{terminals}, respectively,
$R \subseteq N \times (N \cup T)^*$ is a finite set of \emph{production rules},
and $S\in N$ is the \emph{start symbol}.
A rule in $R$ is written as $A \rightarrow \alpha$.
A \emph{stochastic context-free grammar} (SCFG) is a tuple $G=(N,T,R,S,W)$ such that
$(N,T,R,S)$ is a CFG and $W: R \rightarrow[0,1]$ is a function satisfying 
$\sum_{(A \rightarrow \alpha) \in R} W(A \rightarrow \alpha) = 1$ for all $A \in N$.
For every $A \in N$, 
$W$ represents a probability distribution over the set of rules with left-hand side $A$.

\paragraph{Earley Parser}
The Earley Parsing algorithm~\cite{Earley1970} is able to process any SCFG and 
efficiently determine whether a string belongs to the language of the grammar.
We use the Earley parser implementations by~\cite{Trompper2017,Wild2010}
when comparing various SCFGs since it does not require a rigid normal form for grammars.

\paragraph{RNA secondary-structure prediction}
A stochastic context-free grammar can be used for RNA secondary-structure prediction 
where terminals correspond to bases 
and the leftmost derivation of an RNA sequence encodes a secondary structure of the sequence.
The used SCFGs allow many different derivations (and hence secondary structures) for
a given sequence and the rule probabilities induce a probability distribution over those.
Using a classical machine-learning approach, the rule probabilities are chosen as maximum likelihood 
parameters \wrt a given training dataset (with known secondary structures).
For predicting/inferring the (unknown) secondary structure of a new RNA sequence,
a probabilistic parser determines the maximum-likelihood derivation (Viterbi parse)
of the RNA sequence in the SCFG, which encodes the most likely secondary structure
(under the given probabilistic model).

We measure the quality of prediction by \emph{sensitivity} and \emph{positive predictive value (PPV)}:
the fraction of correctly predicted base pairs among all pairs in the reference structure resp.\ all pairs in the predicted structure.
More formally,
let $\mathit{TP}$, $\mathit{TN}$, $\mathit{FP}$, $\mathit{FN}$ be the number of base pairs that are true positives, true negatives, false positives, and false negatives, respectively.
Then $\text{Sensitivity} = \frac{\mathit{TP}}{\mathit{TP}+\mathit{FN}}$ and $\text{PPV}=\frac{\mathit{TP}}{\mathit{TP}+\mathit{FP}}$.

\section{RNA compression using stochastic context-free grammars}
\label{sec:rna-compression-scfg}

We now show how to jointly compress an RNA sequence and secondary structure using a SCFG $G$.
This method has been used by Liu et al.~\cite{LiuYangChenBuZhangYe2008} on a fixed grammar;
we generalize it here to arbitrary grammars $G$ and rule-probability models.
The terminals of $G$ are pairs of characters,
\eg, {\small$\ao$} for base \texttt{A} in the RNA sequence
and \sso in the (dot-bracket representation of the) secondary structure.%
\footnote{%
	Liu et al.\ use 2 grammars instead~-- one for the sequence and one for the secondary structure~--
	the two descriptions are equivalent.
}
To encode an RNA, we determine the sequence of rules in a leftmost derivation of the RNA
and then encode this sequence of rules using a model for the rule probabilities
using a standard code; Liu et al.\ use a fixed Huffman code; we employ arithmetic coding~\cite{Witten1987}.

We illustrate the process on the RNA sequence {\small$\go\au\cc$}
with the grammar of Liu et al.: $G_L = (N,T,R,S)$ has
$N = \{S, L\}$,
$T = \{${\small$\splitaftercomma{ \ao,\co,\go,\uo,  \ac,\cc,\gc,\uc,  \au,\cu,\gu,\uu }$}$\}$,
and rules $R$ shown in \wref{tab:rules_probabilities}.
	\begin{table}[tbh]
	\centering
	\adjustbox{max width=.95\linewidth}{
		\smaller[1]%
		\setlength{\extrarowheight}{5pt}%
		\begin{tabular}{*{3}{|l@{\;\;\;}c@{\;\;\;}c}|} \hline
			\textbf{rule} & \textbf{prob\rlap.} & \textbf{interval}
			& \textbf{rule} & \textbf{prob\rlap.} & \textbf{interval}
			& \textbf{rule} & \textbf{prob\rlap.} & \textbf{interval}\\[3pt] 
			\hline
			$S\to LS$ & 0.65 & $[0.00,0.65)$ &
			$L\to \co S \gc$ & 0.10 & $[0.20,0.30)$ &
			$L\to \au$ & 0.10 & $[0.50,0.60)$ \\
			$S\to \varepsilon$ & 0.35 & $[0.65,1.00)$ &
			$L\to \go S \cc$ & 0.05 & $[0.30,0.35)$ &
			$L\to \uu$ & 0.15 & $[0.60,0.75)$  \\
			$L\to \ao S \uc$ & 0.05 & $[0.00,0.05)$ &
			$L\to \uo S \gc$ & 0.05 & $[0.35,0.40)$ &
			$L\to \cu$ & 0.10 & $[0.75,0.85)$  \\
			$L\to \uo S \ac$ & 0.15 & $[0.05,0.20)$ &
			$L\to \go S \uc$ & 0.10 & $[0.40,0.50)$ &
			$L\to \gu$ & 0.15 & $[0,85,1.00)$ \\[5pt]
			\hline
		\end{tabular}
	}
	\caption{%
		A (fictitious) set of rule probabilities for the grammar of Liu et al.~\cite{LiuYangChenBuZhangYe2008}, including the partition of the unit interval as used in arithmetic coding.
	}
	\label{tab:rules_probabilities}
\end{table}
The (unique) leftmost derivation using the grammar is as follows:
\begin{small}
    \[
        S \Rightarrow 
        LS \Rightarrow
        \go S\cc S \Rightarrow 
        \go LS \cc S \Rightarrow 
        \go \au S \cc S \Rightarrow 
        \go \au\varepsilon\cc S \Rightarrow 
        \go \au \cc\varepsilon = 
        \go \au \cc,
    \]
\end{small}
where the sequence on applied production rules is
\begin{small}
    \[
		S \rightarrow LS, \quad
		L\rightarrow \go S \cc, \quad
		S \rightarrow LS, \quad
		L \rightarrow \au, \quad
		S \rightarrow \varepsilon,\quad
		S \rightarrow \varepsilon.
    \]
\end{small}
Since we always replace the leftmost nonterminal, the next nonterminal to replace is known inductively,
and we can reconstruct the leftmost derivation from only the (index of the) used right-hand sides:
\(
	1,       
	4,      
	1,     
	7,      
	2,     
	2,
\)
using the order of rules in \wref{tab:rules_probabilities}; 
(the $4$ indicates that the second used rule, 
where we know it expends $L$,
is the 4th rule with left-hand side $L$, \ie, $L\to ${\small$\go$}$S${\small$\cc$}).
Now suppose we have the (static) rule probabilities for $R$ from \wref{tab:rules_probabilities}
and we use arithmetic coding to store the right-hand sides.
We obtain the corresponding sequence of intervals from the rules,
$	[0.00,0.65),
	[0.30,0.35),
	[0.00,0.65),
	[0.50,0.60),
	[0.65,1.00),
	[0.65,1.00);
$
which we encode using arithmetic coding to obtain the final binary codeword: \texttt{0011010100100}.

The example above (and~\cite{LiuYangChenBuZhangYe2008}) uses a \emph{static rule-probability model},
usually obtained from a training dataset with known structures by counting how often each rule is used
in the dataset derivations.
With arithmetic coding, we can easily replace this by an \emph{adaptive rule-probability model},
where rule probabilities are computed as relative frequencies in the prefix encoded so far
(starting with some initial value for counters, typically 1). 
This entirely avoids the need for a second pass or a training dataset, as well as storing the rule probabilities.
For long inputs, the adaptive model converges to the sequence-specific relative rule frequencies;
we hence also include the \emph{semi-adaptive model} where rule counts are determined for the given sequence
in a first pass. Unless one also stores the rule counts, this model does not allow decoding, 
but indicates the limiting behavior of the adaptive model.

\section{Joint compression of RNA sequence and secondary structure}
\label{sec:compression}

To investigate the effectiveness of different parameters, we have developed a 
generic prototype implementation in Java that allows us to combine arbitrary SCFGs, 
rule-probability models, and final encoders (Huffman or arithmetic coding).
We use an existing open-source Earley Parser implementation~\cite{Trompper2017}
for obtaining a parse tree (given a SCFG and an RNA with sequence and structure).%
\footnote{%
	This parser has been reported to yield incorrect results for certain inputs;
	for the compression experiment, we could confirm that it works correctly on all our inputs
	and grammars.
} 
Apart from $G_L$ from~\cite{LiuYangChenBuZhangYe2008}, we use the structure-prediction grammars
from~\cite{DowellEddy2004} and~\cite{NebelScheid2011}.
Since non-canonical bonds are regularly found in experimentally determined secondary structures,
all our grammars come in two versions: one that only allows the Watson-Crick and ``\texttt{G}-\texttt{U} wobble'' pairs, and one that allows all 16 pairs.
The difference for compression is small: while most RNA structures do contain non-canonical bonds, most contain only very few of them.

For the compression-quality study, we use the ``friemel'' dataset, consisting of 17\,000 ribosomal RNAs from~\cite{CannoneSubramanian+2002} where
ambiguously sequenced bases, non-canonical base pairs and pseudoknots have been removed~\cite{Friemel2020}.
Information of each RNA in the given datasets is stored in a text file, using the dot-bracket notation. 
24 contained empty hairpin loops; since 2 grammars from~\cite{DowellEddy2004} exclude these, we replaced the innermost pair by two unpaired bases; for the evaluation, we exclude these 24 RNAs.

\begin{figure}[tbhp]
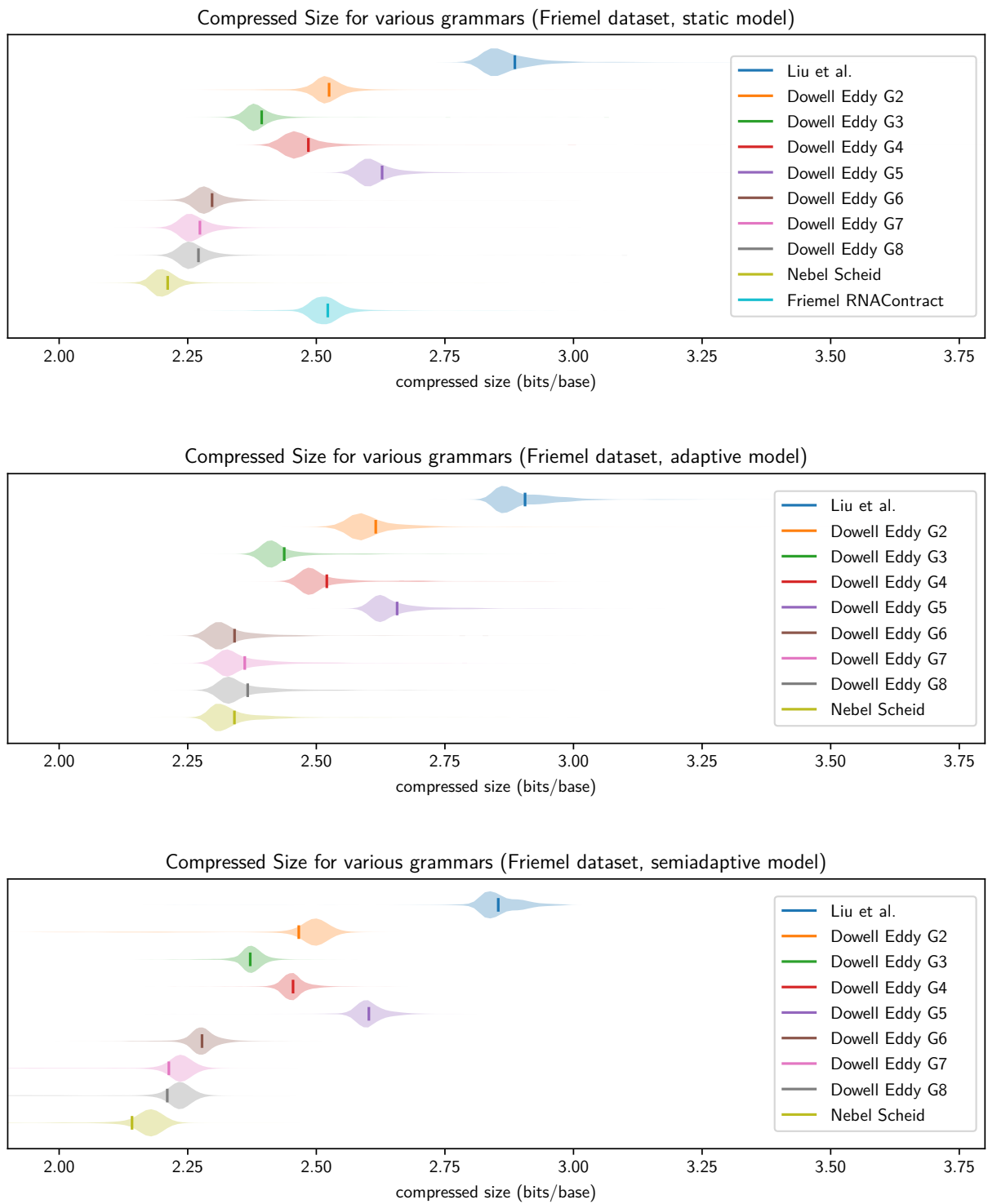

\ifdcc{%
	\def\myscale{0.5}%
}{%
	\def\myscale{0.8}
}
	\plaincenter{\scalebox{\myscale}{\input{pics/compressed-sizes-friemel-static.pgf}}}\\
	\plaincenter{\scalebox{\myscale}{\input{pics/compressed-sizes-friemel-adaptive.pgf}}}\\
	\plaincenter{\scalebox{\myscale}{\input{pics/compressed-sizes-friemel-semiadaptive.pgf}}}
	\caption{%
		Means (vertical bars) and distributions (shaded violin plot) of the normalized compressed 
		size using various grammars on Friemel's RNA dataset.
		All compressed sizes are shown as bits per base.
		\textbf{Top:} results using static rule probabilities, determined from the entire dataset.
		\textbf{Middle:} results using adaptive rule-probabilities model (LaPlace model).
		\textbf{Bottom:} semi-adaptive rule probabilities (ignoring space for storing rule probabilities).
	}
	\label{fig:compressed-size-friemel}
\end{figure}

\wref{fig:compressed-size-friemel} shows the compression quality of different grammars,
normalized to the (average) number of bits per base in the RNA.
It is striking that the current state-of-the-art method from the literature, Liu et al.'s RNACompress~\cite{LiuYangChenBuZhangYe2008}, performs much worse than all the structure-prediction grammars
(for all rule-probability models),
indicating that these grammars indeed incorporate effective domain knowledge about RNA structures.
Also note that a simplistic encoding of the RNA sequence alone would use 2 bits/base;
the most sophisticated grammars come very close to that for the joint encoding of sequence \emph{and} structure: 2.21 bits/base on average for the grammar of Nebel and Scheid~\cite{NebelScheid2011}.
The large grammars $G_2$, $G_7$, and $G_8$~\cite{DowellEddy2004} (those with ``stacking parameters'') 
and the huge grammar by Nebel and Scheid~\cite{NebelScheid2011}
perform overall best. 
But some much smaller grammars like $G_6$ come very close, despite having a factor 10 fewer parameters. 
This shows that it is the structure of the grammar, not merely the number of parameters of the model,
that improve compression of RNA secondary structures.

\ifdraft{\FloatBarrier}

\section{Compression ratio vs.\ prediction quality}
\label{sec:compression-vs-preduction}

We have seen that the choice of the grammar heavily influences the compression quality 
of our generic joint RNA compressor.
In this section, we take a closer look at this grammar dependence from 
the perspective of both compression and secondary-structure prediction.
For that, we reproduced the classic study of Dowell and Eddy~\cite{DowellEddy2004}
comparing several hand-crafted SCFG for their ability to correctly infer RNA secondary structures
given only the RNA sequence as input.
Due to the bugs from~\cite{Trompper2017}, we here used the probabilistic Earley parser from~\cite{Wild2010}.
We use the original datasets from~\cite{DowellEddy2004} (available at \url{http://eddylab.org/software/conus/}):
The ``benchmark'' dataset was used in~\cite{DowellEddy2004} to compare the prediction quality of SCFGs,
whose rule probabilities have been trained on their ``mixed80'' dataset; see~\cite{DowellEddy2004} for further details.
Both datasets contain many non-canonical bonds and 8 RNAs contain empty hairpin loops; we again eliminated the latter. 
Mixed80 contains numerous ambiguous bases; these were randomly replaced with a compatible base.

\begin{figure}[tbh]
	\small

    \ifdcc{
    	\begin{minipage}[t]{.55\textwidth}
    		~\\[-1ex]
    		{\scalebox{.53}{\large\input{pics/compression-vs-prediction-static-mixed80.pgf}}}
    	\end{minipage}%
    	\begin{minipage}[t]{.45\textwidth}
    		\caption{%
    		Scatter plot of compression vs.\ prediction quality for the grammars from~\cite{DowellEddy2004}.
    		Each grammar is presented as one point with error bars.
    		The $x$-axis shows the compressed size (in bits per base) for joint compression of 
    		RNA sequence and secondary structure, averaged over the benchmark dataset~\cite{DowellEddy2004}.
    		Horizontal error bars show one standard deviation of compressed size over the benchmark dataset.
            The $y$-axis shows the geometric\linebreak%
            \label{fig:compression-prediction-static-mixed80}
    	}
    	\end{minipage}
    	~\\[-2.1\baselineskip]
             mean of sensitivity and PPV (for each predicted RNA secondary structure, 
            averaged over the benchmark dataset); error bars show one standard deviation.
            For the ambiguous grammars $G_1$ and $G_2$, no vertical error bars are available (we did not reproduce predictions for these; the average is taken from~\cite{DowellEddy2004}).
    		Both compression and prediction use the same training dataset (mixed80 from~\cite{DowellEddy2004})
    		to determine the parameters of the grammars; compression here uses the static model for rule probabilities.
	}{
        \plaincenter{\scalebox{.75}{\large\input{pics/compression-vs-prediction-static-mixed80.pgf}}}
        \caption{%
    		Scatter plot of compression vs.\ prediction quality for the grammars from~\cite{DowellEddy2004}.
    		Each grammar is presented as one point with error bars.
    		The $x$-axis shows the compressed size (in bits per base) for joint compression of 
    		RNA sequence and secondary structure, averaged over the benchmark dataset~\cite{DowellEddy2004}.
    		Horizontal error bars show one standard deviation of compressed size over the benchmark dataset.
            The $y$-axis shows the geometric
             mean of sensitivity and PPV (for each predicted RNA secondary structure, 
            averaged over the benchmark dataset); error bars show one standard deviation.
            For the ambiguous grammars $G_1$ and $G_2$, no vertical error bars are available (we did not reproduce predictions for these; the average is taken from~\cite{DowellEddy2004}).
    		Both compression and prediction use the same training dataset (mixed80 from~\cite{DowellEddy2004})
    		to determine the parameters of the grammars; compression here uses the static model for rule probabilities.
        }
    }
   \ifdcc{}{\label{fig:compression-prediction-static-mixed80}}
\end{figure}

\wref{fig:compression-prediction-static-mixed80} shows the results of comparing for each grammar how well it compresses the benchmark dataset of RNAs and how well it predicts secondary structures of this set (using the setup and parameters as in~\cite{DowellEddy2004}).
Taking into account the variability across different RNAs within the dataset,
a clear and strong negative correlation is visible between compressed size and prediction quality;
in particular, there is a clearly distinct cluster of grammars that simultaneously give the best
compression and the best prediction.
\begin{figure}[tbp]
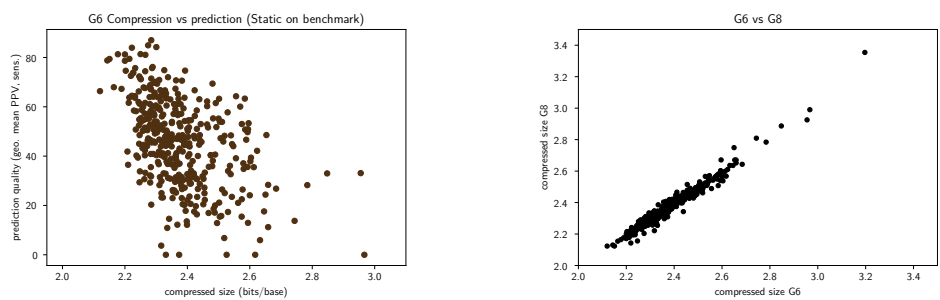

	\plaincenter{%
		\scalebox{.4}{\large\input{pics/compression-vs-prediction-G6-static-mixed80.pgf}}\hfill%
		\scalebox{.4}{\large\input{pics/compressed-size-benchmark-G6-vs-G8.pgf}}
	}
	\caption{%
		Scatter plots with one point per RNA sequence in the benchmark dataset.
		\textbf{Left:} compressed size against prediction quality using $G_6$.
		\textbf{Right:} compressed size using $G_6$ against compressed size using $G_8$.
		All compression methods use the static rule probabilities trained on mixed80.
	}
	\label{fig:scatter-plots-per-rna}
\end{figure}
At least for the grammars from~\cite{DowellEddy2004}, this shows that one can use compressed size as a
more rigidly defined and robust proxy for secondary-structure prediction quality.

\wref{fig:scatter-plots-per-rna} takes a closer look at the correlation on a per-RNA level.
Even there, a correlation remains visible; in particular very accurately predicted structures are also well compressed.
The right panel in \wref{fig:scatter-plots-per-rna} shows that compressed size for different grammars is 
very strongly correlated; pictures for other grammar pairs are similar (excluding the poor performing $G_1$, $G_4$, and $G_5$).
Note that despite the strong correlation at RNA level, there is a significant difference in the (mean) compression ratio between different grammars.
This might indicate that there are intrinsically more and less ``surprising'' RNA secondary structures (knowing only the RNA sequence).

\section{Conclusion}
\label{sec:conclusion}

In this paper, we demonstrated how domain knowledge of RNA secondary structures 
encapsulated in stochastic context-free grammars for structure prediction
can be used to obtain the best single-RNA compression ratios known for this type of data.
Moreover, we showed promising first evidence for the utility of compression ability
as a cheap and robust proxy for prediction quality for RNA secondary-structure prediction.

This work opens up several enticing avenues for future research.
Using compression ability as simpler guide, we are working on an approach 
to discover new promising models for secondary-structure prediction.
It would be interesting to investigate whether the robust correlation between prediction quality
and compressed size continues to hold for large grammars with many parameters;
here prediction could suffer due to overfitting issues, 
whereas compression might continue see improvements from additional parameters.
Since many natural RNA secondary structures contain ``pseudoknots'', a principled approach
for compressing such structures would be interesting.
If the compression-prediction correlation can be demonstrated in this domain as well,
the lack of reliably free-energy models for pseudoknotted RNA structures and the relative lack 
of high-fidelity training data would make compression ability of even greater value 
in the search for better predictions models.

	\myacknowledgements
%

%
\ifdcc{\section*{References}}{}
\begin{small}
\bibliography{references}

\begin{thebibliography}{10}

\bibitem{CannoneSubramanian+2002}
J.~J. Cannone~et al.
\newblock The {Comparative RNA Web (CRW)} site: an online database of
  comparative sequence and structure information for ribosomal, intron, and
  other {RNAs}.
\newblock {\em BMC Bioinformatics}, 3, 2002.

\bibitem{RNACentral2020}
R.~Consortium.
\newblock {RNAcentral} 2021: secondary structure integration, improved sequence
  search and new member databases.
\newblock {\em Nucleic Acids Research}, 49(D1):D212--D220, 2020.

\bibitem{DowellEddy2004}
R.~D. Dowell and S.~R. Eddy.
\newblock Evaluation of several lightweight stochastic context-free grammars
  for rna secondary structure prediction.
\newblock {\em BMC bioinformatics}, 5(1):1--14, 2004.

\bibitem{DurbinEddyKroghMitchison1998}
R.~Durbin, S.~R. Eddy, A.~Krogh, and G.~Mitchison.
\newblock {\em Biological sequence analysis: probabilistic models of proteins
  and nucleic acids}.
\newblock Cambridge university press, 1998.

\bibitem{Earley1970}
J.~Earley.
\newblock An efficient context-free parsing algorithm.
\newblock {\em Communications of the ACM}, 13, 1970.

\bibitem{Friemel2020}
J.~Friemel.
\newblock {\em Contraction-Based Compression of {RNA} Secondary Structures}.
\newblock {BSc} dissertation, Universitat Bielefeld, 2020.

\bibitem{FuCaoWuPengNieXie2021}
L.~Fu, Y.~Cao, J.~Wu, Q.~Peng, Q.~Nie, and X.~Xie.
\newblock {UFold}: fast and accurate {RNA} secondary structure prediction with
  deep learning.
\newblock {\em Nucleic Acids Research}, 50(3):e14--e14, 2021.

\bibitem{GiancarloScaturroUtro2009}
R.~Giancarlo, D.~Scaturro, and F.~Utro.
\newblock Textual data compression in computational biology: a synopsis.
\newblock {\em Bioinformatics}, 25, 2009.

\bibitem{GorodkinRuzzo2014}
J.~Gorodkin and W.~L. Ruzzo, editors.
\newblock {\em {RNA} Sequence, Structure, and Function: Computational and
  Bioinformatic Methods}.
\newblock Humana Press, 2014.

\bibitem{Hofacker1994}
I.~Hofacker, W.~Fontana, P.~Stadler, L.~Bonhoeffer, M.~Tacker, and P.~Schuster.
\newblock Fast folding and comparison of {RNA} secondary structures.
\newblock {\em Chemical monthly}, 125:167--168, 1994.

\bibitem{KalvariEtAl2020}
I.~Kalvari~et al.
\newblock Rfam 14: expanded coverage of metagenomic, viral and {microRNA}
  families.
\newblock {\em Nucleic Acids Research}, 49(D1):D192--D200, 2020.

\bibitem{KiefferYang2022}
J.~C. Kieffer and E.~hui Yang.
\newblock Survey of grammar-based data structure compression.
\newblock {\em {IEEE} {BITS} the Information Theory Magazine}, pages 1--12,
  2022.

\bibitem{Kolmogorov1998}
A.~Kolmogorov.
\newblock On tables of random numbers.
\newblock {\em Theoretical Computer Science}, 207, 1998.

\bibitem{LarssonMoffat2000}
N.~Larsson and A.~Moffat.
\newblock Off-line dictionary-based compression.
\newblock {\em Proceedings of the {IEEE}}, 88(11):1722--1732, 2000.

\bibitem{LeeEtAl2014}
J.~Lee~et al.
\newblock {RNA} design rules from a massive open laboratory.
\newblock {\em Proceedings of the National Academy of Sciences},
  111(6):2122--2127, 2014.

\bibitem{LiuYangChenBuZhangYe2008}
Q.~Liu, Y.~Yang, C.~Chen, J.~Bu, Y.~Zhang, and X.~Ye.
\newblock {RNACompress}: Grammar-based compression and informational complexity
  measurement of {RNA} secondary structure.
\newblock {\em BMC bioinformatics}, 9(1):1--12, 2008.

\bibitem{Maneth2018}
S.~Maneth.
\newblock Grammar-based compression.
\newblock In {\em Encyclopedia of Big Data Technologies}, pages 1--8. Springer
  International Publishing, 2018.

\bibitem{Mathews2019}
D.~H. Mathews.
\newblock How to benchmark rna secondary structure prediction accuracy.
\newblock {\em Methods}, 162:60--67, 2019.

\bibitem{NaganumaHendrianYoshinakaShinoharaKobayashi2020}
H.~Naganuma, D.~Hendrian, R.~Yoshinaka, A.~Shinohara, and N.~Kobayashi.
\newblock Grammar compression with probabilistic context-free grammar.
\newblock In {\em 2020 Data Compression Conference ({DCC})}. {IEEE}, 2020.

\bibitem{NebelScheid2011}
M.~E. Nebel and A.~Scheid.
\newblock Evaluation of a sophisticated {SCFG} design for {RNA} secondary
  structure prediction.
\newblock {\em Theory in Biosciences}, 130(4):313--336, 2011.

\bibitem{RivasLangEddy2011}
E.~Rivas, R.~Lang, and S.~R. Eddy.
\newblock A range of complex probabilistic models for {RNA} secondary structure
  prediction that includes the nearest-neighbor model and more.
\newblock {\em {RNA}}, 18(2):193--212, 2011.

\bibitem{SatoAkiyamaSakakibara2021}
K.~Sato, M.~Akiyama, and Y.~Sakakibara.
\newblock {RNA} secondary structure prediction using deep learning with
  thermodynamic integration.
\newblock {\em Nature Communications}, 12(1), 2021.

\bibitem{SatoKato2021}
K.~Sato and Y.~Kato.
\newblock Prediction of {RNA} secondary structure including pseudoknots for
  long sequences.
\newblock {\em Briefings in Bioinformatics}, 23(1), 2021.

\bibitem{Schulz2012}
A.~Schulz.
\newblock {\em Sampling and Approximation in the Context of RNA Secondary
  Structure Prediction Algorithms and Studies Based on Stochastic Context-Free
  Modeling}.
\newblock {PhD} dissertation, Technische Universität Kaiserslautern, 2012.

\bibitem{Trompper2017}
M.~Trompper~{(digitalheir)}.
\newblock Probabilistic earley parser, 2017.

\bibitem{TurnerMathews2016}
D.~H. Turner and D.~H. Mathews, editors.
\newblock {\em {RNA} Structure Determination}.
\newblock Springer New York, 2016.

\bibitem{Wild2010}
S.~Wild.
\newblock {\em An Earley-style parser for solving the RNA-RNA interaction
  problem.}
\newblock {BSc} dissertation, University of Kaiserslautern, 2010.

\bibitem{Witten1987}
I.~H. Witten, R.~M. Neal, and J.~G. Cleary.
\newblock Arithmetic coding for data compression.
\newblock {\em Communications of the ACM}, 30, 1987.

\end{thebibliography}
\end{small}

	\clearpage
	\appendix
	\ifkoma{\addpart{Appendix}}{}
	
	\section{Comparison with general purpose compressors}
\label{app:paq8l}

To compare the compression quality of our approach with state-of-the-art generic compressors, 
we use the \emph{paq8l} tool (\url{http://mattmahoney.net/dc/#paq}).
We compressed each individual RNA text file (with sequence in the first line and the secondary structure as dot-bracket string in the second line) in the \texttt{friemel-modified} dataset using \texttt{paq8l -8} (the setting for best compression) and summed up the file sizes of all compressed RNAs.

The uncompressed size of \texttt{friemel-modified} is 39\,284\,962 bytes and 
all RNAs combined have 19\,357\,501 bases (2 bytes per base, one for sequence, one for structure, plus a small amount of metadata overhead).
paq8l compressed this to 9\,146\,548 bytes. Dividing this total compressed size (in bytes) by the total number of bases in the dataset yields an average of 3.78 bits per base.
This is 70\% more than the 2.211 bits that our compressed with $G_S$ achieves (using a static rule-probability model).

It is not unexpected that a general purpose tool like paq8l does not
come anywhere close to the compression of a domain-aware model;
however, it is a bit surprising that paq8l uses substantially more space than the 
local first order empirical entropy: All first lines of the files have letters in $\{\texttt{A}, \texttt{C}, \texttt{G}, \texttt{U}\}$, and thus a local entropy of at most 2 bits per character.
For the second line, we only have $\{\sso,\ssc,\ssu\}$, and hence at most $\lg(3)\approx 1.58$ bits per character.
Exploiting this local entropy would result in 3.58 bits per base.
	
	\section{Grammars}
\label{app:grammars}
{
\setlength{\parindent}{0pt}
\setlength{\parskip}{\medskipamount}

Here, we list the used grammars; we use the compact notation from~\cite{DowellEddy2004},
where we only give $a$ and $\hat a$ as terminals instead of the pairs introduced in \wref{sec:preliminaries}. The actual RNA grammars
would have 4 rules for each rule with a single ``$a$''; instead of $A\to \alpha a \beta$, we would actually have
$A\to \alpha \au \beta$,
$A\to \alpha \cu \beta$,
$A\to \alpha \gu \beta$, and
$A\to \alpha \uu \beta$;
similarly, each rules with a ``$a \hat a$'' pair actually stands for 6 rules resp.\ 16 rules if we allow non-canonical base pairs.
For the stacking grammars, nonterminals $B^{a\hat a}$ are shorthand notation for 6 resp.\ 16 different nonterminals, which ``remember'' an enclosing pair. 
If there are several occurrences of the same $a\hat a$ pair within one rule, these must be replaced
consistently (with the same bases in all occurrences).

Our parsers require grammars to be free of $\epsilon$-rules, 
so we eliminated these in all grammars.

Moreover, the fast stochastic parser used for the prediction study requires a slightly more restrictive form:
the grammars are not allowed to have left-recursive rules,
and the nonterminals must be ordered, so that $B$ comes before $A$ whenever one can derive $B\alpha$ from $A$.
We only use the unambiguous grammars $G_3,\ldots,G_8$ for the prediction study,
so we directly give those grammars in the required form.

\subsection*{\boldmath Grammar $G_{L'}$ (\texttt{LiuGrammar})}

The first grammar is $G_{L'}$ from Liu et al.~\cite{LiuYangChenBuZhangYe2008}
where we eliminate $\epsilon$-rules.

$T \to a$ \quad (4 rules) \\
$T \to a S \hat a$ \quad (16 rules) \\
$S \to T \mid TS$

\subsection*{\boldmath Grammar $G_1$ (\texttt{DowellGrammar1Bound})}

Next, 
$G_1,\ldots,G_8$ are the grammars taken from Dowell and Eddy~\cite{DowellEddy2004}.

$U \to a$                                   \\
$B \to a S \hat a$                           \\
$C \to B \mid U$			\\
$X \to UX \mid SX \mid U \mid S$		\\
$S \to C \mid CX \mid US \mid USX$

\subsection*{\boldmath Grammar $G_2$ (\texttt{DowellGrammar2Bound})}

$U\to a$                                           \\
$P^{a\hat a} \to a P^{a\hat a} \hat a \mid S$               \\
$S \to a P^{a\hat a} \hat a \mid U \mid US \mid SU \mid SS$

\subsection*{\boldmath Grammar $G_3$ (\texttt{DowellGrammar3Bound})}

$U\to a$                                           \\
$B\to a S \hat a$                                   \\
$L \to B \mid U L$                                   \\
$R \to U \mid U R$                                   \\
$S \to B \mid UL \mid RU \mid LS \mid U$

\subsection*{\boldmath Grammar $G_4$ (\texttt{DowellGrammar4Bound})}

$U\to a$                                           \\
$B\to a S \hat a$                                   \\
$C \to B \mid U$                                   \\
$D \to C \mid CD$                                   \\
$Q \to B \mid BD$                                   \\
$S \to U \mid US \mid Q$

\subsection*{\boldmath Grammar $G_5$ (\texttt{DowellGrammar5Bound})}

$U\to a$                                            \\
$B\to a S \hat a$                                   \\
$S \to U \mid B \mid US\mid BS$

\subsection*{\boldmath Grammar $G_6$ (\texttt{DowellGrammar6Bound})}

$U \to a$          \\
$B \to a M \hat a $   \\
$T \to B \mid U$      \\
$M \to B \mid TS \mid T$     \\
$S\to TS\mid T$       

An alternative version does not have the rule $M\to T$; 
that grammar then disallows hairpins of length one, \ie, `\sso\ssu\ssc`.

\subsection*{\boldmath Grammar $G_7$ (\texttt{DowellGrammar7Bound})}

$U \to a$ \\
$B\to a V^{a\hat a} \hat a$   \quad (16 rules) \\
$B^{b \hat b} \to a V^{a\hat a} \hat a$   \quad ($16\cdot 16$ rules) \\
$L \to B \mid UL$                            \\
$M \to UM \mid U$                            \\
$T \to U \mid UL \mid MU \mid LS$            \\
$V^{a \hat a} \to B^{a\hat a} \mid T$   \quad ($16\cdot 2$ rules) \\
$S \to B \mid UL \mid MU \mid U \mid LS$     
	
\subsection*{\boldmath Grammar $G_8$ (\texttt{DowellGrammar8Bound})}

$U \to a$ \\
$B \to a V^{a\hat a} \hat a$                \\
$B^{b \hat b} \to a V^{a\hat a} \hat a$    \\
$C \to U \mid B$ \\
$D \to C \mid CD$  \\
$E \to B \mid BD$                  \\
$N \to U \mid E \mid US \mid EU \mid EB$          \\
$V^{a \hat a} \to B^{a\hat a} \mid N$      \\
$S \to U \mid E \mid US$

\subsection*{\boldmath Grammar $G_S$ (\texttt{SchulzGrammar})}

The grammar $G_S$ is taken from~\cite{NebelScheid2011}; see also~\cite[Def.\,A.1.2]{Schulz2012};
we have made the modifications described below to make the grammar more suitable for compression.

\newcommand\ub{\ssu}
\newcommand\ubE{\ensuremath{a}\xspace}
\newcommand\ubN[1]{\ensuremath{X^{#1}}\xspace}
\newcommand\uN[1]{\ensuremath{X^{#1}}\xspace}
\newcommand\po{\ensuremath{a}\xspace}
\newcommand\pc{\ensuremath{\hat a}\xspace}

Since we have to expand every occurrence of \po\pc on the right-hand side into 6 (or even 16) rules 
in our RNA grammars, we replaced ``$\po L\pc $'' in several right-hand sides with a nonterminal that expands to $\po L\pc$ ($A$ when we start a new stem and the new nonterminal $I$ when we continue after an interior loop or bulge).
This reduces the number of parameters and hence the expressive power a bit, but will keep the grammar substantially smaller.

\def\myskip{\quad\linebreak[2]}

$p_{0}':S'\rightarrow S,$

$p_{1}':S\rightarrow A,\myskip $
$p_{2}':S\rightarrow AC,\myskip 
p_{3}':S\rightarrow TA,\myskip 
p_{4}':S\rightarrow TAC,$

$p_{5}':T\rightarrow A,\myskip 
p_{6}':T\rightarrow AC,\myskip p_{7}':T\rightarrow TA,\myskip p_{8}':T\rightarrow TAC,$

$p_{9}':T\rightarrow C,$

$p_{10}':C\rightarrow\ubN{C} ,\myskip p_{11}':C\rightarrow C\ubN{C} ,$

$p_{12}':A\rightarrow \po L \pc,$

$p_{13}':L\rightarrow \po L\pc,\myskip p_{14}':L\rightarrow M,\myskip p_{15}':L\rightarrow P,\myskip p_{16}':L\rightarrow Q,$

$p_{17}':L\rightarrow R,\myskip p_{18}':L\rightarrow F,\myskip p_{19}':L\rightarrow G,$

$p_{20}':G\rightarrow I  \ubE ,\myskip 
p_{21}':G\rightarrow I   \ubN{B}\ubN{B} ,\myskip 
p_{22}':G\rightarrow I  B \ubN{B}\ubN{B} ,$

$p_{23}':G\rightarrow\ubE  I  \myskip 
p_{24}':G\rightarrow \ubN{B}\ubN{B}  I  \myskip 
p_{25}':G\rightarrow \ubN{B}\ubN{B} B I  $

$p_{26}':B\rightarrow \ubN{B} \myskip p_{27}':B\rightarrow B\ubN{B} $

$p_{28}':F\rightarrow \ubN{F}\ubN{F}\ubN{F} \myskip 
p_{29}':F\rightarrow \ubN{F}\ubN{F}\ubN{F}\ubN{F} \myskip 
p_{30}':F\rightarrow \ubN{F}\ubN{F}\ubN{F}\ubN{F}\ubN{F} \\
p_{31}':F\rightarrow \ubN{F}\ubN{F}\ubN{F}\ubN{F}\ubN{F} H$

$p_{32}':H\rightarrow\ubN{H} \myskip 
p_{33}':H\rightarrow H\ubN{H} $

$p_{34}':P\rightarrow\ubE  I  \ubE \myskip 
p_{35}':P\rightarrow\ubN{I}  I   \ubN{I}\ubN{I} \myskip 
p_{36}':P\rightarrow \ubN{I}\ubN{I}  I  \ubN{I} \myskip 
p_{37}':P\rightarrow \ubN{I}\ubN{I}  I   \ubN{I}\ubN{I} $

$p_{38}':Q\rightarrow \ubN{I}\ubN{I}  I   \ubN{I}\ubN{I}\ubN{I} \myskip 
p_{39}':Q\rightarrow \ubN{I} \ubN{I}   I  K \ubN{I} \ubN{I} \ubN{I}  \myskip 
p_{40}':Q\rightarrow \ubN{I} \ubN{I} \ubN{I}   I   \ubN{I} \ubN{I}  \myskip 
p_{41}':Q\rightarrow \ubN{I} \ubN{I} \ubN{I}  J I  \ubN{I}\ubN{I}  $

$p_{42}':Q\rightarrow \ubN{I} \ubN{I} \ubN{I}   I  K \ubN{I} \ubN{I}  \myskip 
p_{43}':Q\rightarrow \ubN{I} \ubN{I} \ubN{I}  J I  K \ubN{I} \ubN{I}  $

$p_{44}':R\rightarrow\ubN{I}   I   \ubN{I} \ubN{I} \ubN{I}  \myskip 
p_{45}':R\rightarrow\ubN{I}   I  K \ubN{I} \ubN{I} \ubN{I}  \myskip 
p_{46}':R\rightarrow \ubN{I} \ubN{I} \ubN{I}   I  \ubN{I}  \myskip 
p_{47}':R\rightarrow \ubN{I} \ubN{I} \ubN{I}  J I  \ubN{I}  $

$p_{48}':J\rightarrow\ubN{I} \myskip p_{49}':J\rightarrow{}J\ubN{I} $

$p_{50}':K\rightarrow\ubN{I} \myskip p_{51}':K\rightarrow{}K\ubN{I} $

$p_{52}':M\rightarrow AA\myskip 
p_{53}':M\rightarrow{}UAA\myskip 
p_{54}':M\rightarrow{}AUA\myskip 
p_{55}':M\rightarrow{}AAN$

$p_{56}':M\rightarrow UAUA\myskip 
p_{57}':M\rightarrow{}UAAN\myskip 
p_{58}':M\rightarrow{}AUAN\myskip 
p_{59}':M\rightarrow{}UAUAN$

$p_{60}':N\rightarrow A$\myskip 
$p_{61}':N\to UA$\myskip 
$p_{62}':N\to AN$\myskip 
$p_{63}':N\to UAN$\myskip 

$p_{64}':N\rightarrow U$

$p_{65}':U\rightarrow \ubN{U} \myskip p_{65}':U\rightarrow U\ubN{U} $

We add the following rules:

$F \to \ubN{F}$\myskip 
$F \to \ubN{F}\ubN{F}$
\myskip  (allow length 1 and 2 in hairpins)

$I \to \po L\pc$
\myskip  (new nonterminal for use inside bulges/interior loops)

$S \to C$\myskip (allow completely unpaired sequences)

Rules for all unpaired nonterminals:\\
$\uN{B} \to \ubE$,\myskip
$\uN{C} \to \ubE$,\myskip
$\uN{F} \to \ubE$,\myskip
$\uN{H} \to \ubE$,\myskip
$\uN{I} \to \ubE$,\myskip
$\uN{U} \to \ubE$

}
	
	\section{Further results}
\label{app:further-results}

This appendix reports on some further results that were left out of the main text
due to space constraints in the proceedings version.

\subsection{Huffman coding vs.\ Arithmetic coding}

We here compare the influence of the coding step on compression ratio in isolation.
For that, we modify Liu et al.'s RNACompress~\cite{LiuYangChenBuZhangYe2008} to use 
arithmetic coding instead of a Huffman code, leaving everything else unchanged, 
and compare the results.

We were not able to obtain the original implementation of RNACompress 
and the datasets from Liu et al.~\cite{LiuYangChenBuZhangYe2008}.
We hence re-implemented  RNACompress, and used the friemel-modified dataset
of 17\,000 RNA samples originally taken from \cite{CannoneSubramanian+2002}
instead of the dataset from~\cite{LiuYangChenBuZhangYe2008}.
Some of the RNAs in Friemel's dataset have non-canonical bonds 
(these are less stable secondary bonds). 
Since Liu et al.\ do not allow non-canonical bonds in their tool, we also removed these from Friemel's dataset,
\ie, we replaced the open \sso and close \ssc parenthesis for non-canonical bonds with unpaired bases \ssu in the positions were non-canonical bonds appeared. 
Afterwards only the stable bonds (Watson-Crick and G--U wobble bonds) were left in all samples in the dataset,
which we call friemel-modified.
    
Unsurprisingly, the arithmetic coding produced better compression results than Huffman coding, but the difference between the means is only 2.7\%. \wref{fig:Comparing_Huffman_and_AcCompressor_violin plot} shows the distribution of compressed size over the RNAs; while arithmetic coding has moderate impact on the mean compressed size, it helps a lot to bring down the right tail.
The scatterplot in \wref{fig:Comparing_Huffman_and_AcCompressor_scatterplot} further shows that indeed,
arithmetic coding (with this fixed static model) is doing better on almost all RNAs, and the effect is bigger for those RNAs that are compressed worse.

\begin{figure}[htbp]
    \centering
    \includegraphics[width=.8\linewidth]{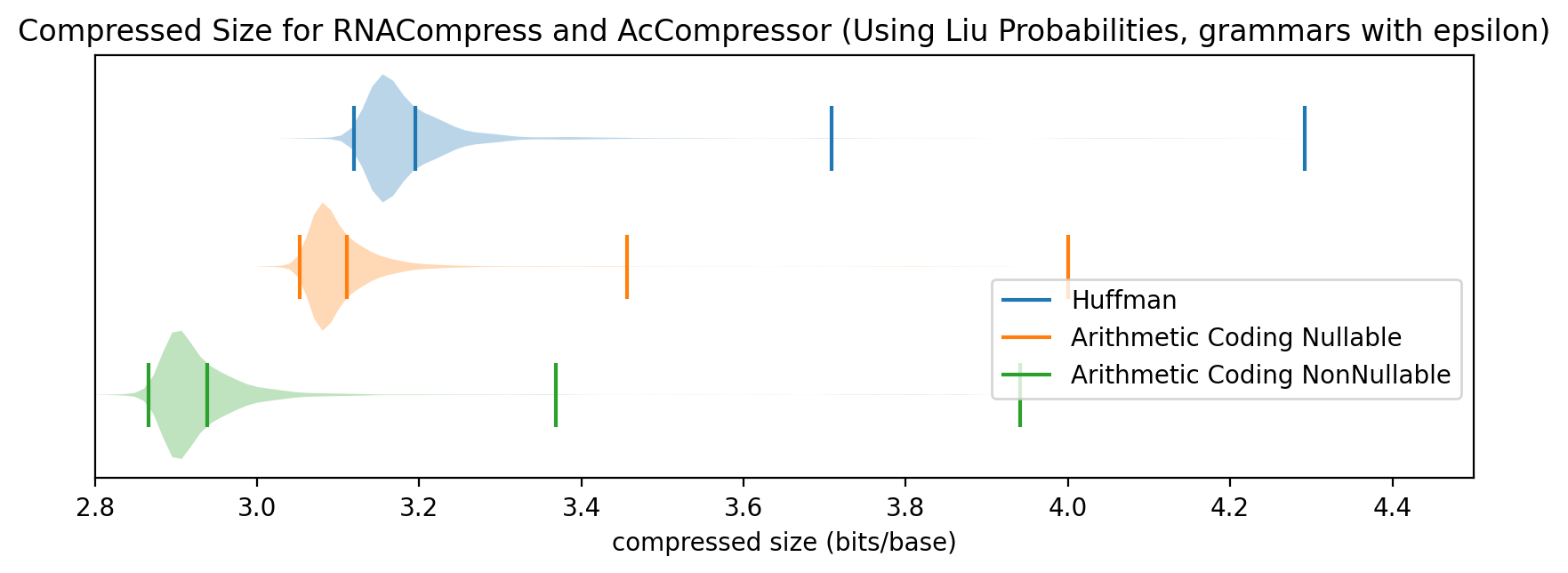}
    \caption{%
       	Compressed size in bits per base for RNACompress (original with Huffman coding) and RNACompress with arithmetic coding, and the RNACompress variant with the $\epsilon$-rule-free grammar.
       	The vertical bars show from, left to right, the 1\% quantile, mean, 99\% quantile, and maximum.
       	The means are at 3.195 resp.\ 3.110 bits per base.
    }
    \label{fig:Comparing_Huffman_and_AcCompressor_violin plot}
\end{figure}

\begin{figure}[htbp]
    \centering
    \includegraphics[width=.48\linewidth]{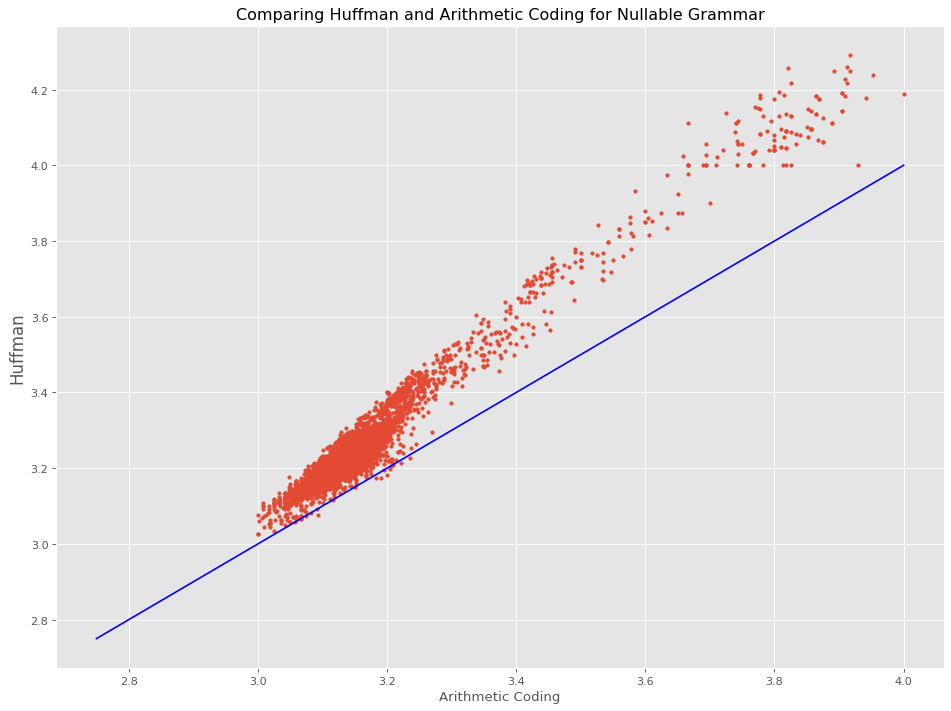}
    \includegraphics[width=.48\linewidth]{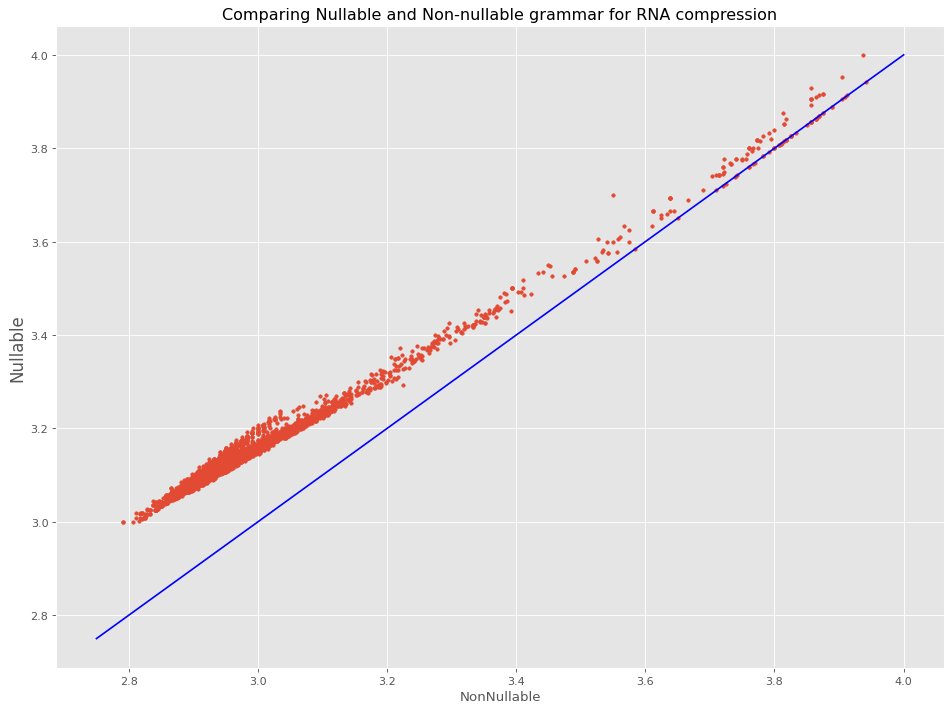}
    \caption{The same data as in \wref{fig:Comparing_Huffman_and_AcCompressor_violin plot}, but as scatter plots with one point per RNA.}
    \label{fig:Comparing_Huffman_and_AcCompressor_scatterplot}
\end{figure}

\subsection{Nullable Grammar vs.\ Non-Nullable Grammar}

Liu et al.~\cite{LiuYangChenBuZhangYe2008} originally use the following grammar
(in the notation from \wref{app:grammars}):

\medskip

\noindent
$G_L$\\
$L \to a S \hat a \mid a$\\
$S \to LS \mid \epsilon$

\medskip

\noindent 
For general parsers, $\epsilon$-rules are often inconvenient;
we therefore modified this grammar to $G_{L'}$ shown in \wref{app:grammars}.
This transformation makes the probabilistic model slightly richer and so will help compression, but 
it does not change the nature of the grammar; the structure of leftmost derivations of strings remain (almost) the same.
(We here ignore the fact that the empty string is no longer in the language of grammar $G_{L'}$, 
while it was derivable in $G_L$. For RNA compression, this is not relevant.)
We manually implemented a parser for the original $G_L$ grammar and compared the compression outcome.
As \wref{fig:Comparing_Huffman_and_AcCompressor_violin plot} shows, this very moderate enrichment of the 
probabilistic model has a larger impact than moving from Huffman to arithmetic coding.
The scatter plot in \wref{fig:Comparing_Huffman_and_AcCompressor_scatterplot} (right) shows that again, we never do worse in $G_{L'}$ compared to $G_L$, but that this time, the biggest savings are happening for the
(much larger number of) RNAs that are compressed \emph{well}.

\end{document}